\documentclass[
12pt,tightenlines, amsmath, amssymb, nofootinbib, prd,
superscriptaddress, showpacs, preprintnumbers]{revtex4}
\newcommand{\beq}{\begin{equation}}
\newcommand{\eeq}{\end{equation}}
\newcommand{\ba}{\begin{array}{ccc}}
\newcommand{\ea}{\end{array}}

\def\beqn{\begin{eqnarray}}
\def\eeqn{\end{eqnarray}}

\def\<{\langle}
\def\>{\rangle}

\def\spose#1{\hbox to 0pt{#1\hss}}
\def\ltapprox{\mathrel{\spose{\lower 3pt\hbox{$\mathchar"218$}}
 \raise 2.0pt\hbox{$\mathchar"13C$}}}
\def\gtapprox{\mathrel{\spose{\lower 3pt\hbox{$\mathchar"218$}}
 \raise 2.0pt\hbox{$\mathchar"13E$}}}
\usepackage{graphicx}
\usepackage{amsmath}
\usepackage{amssymb}
\usepackage{psfrag}
\usepackage{epsfig}
\begin{document}

\preprint{DFUP--TH/2006--9}
\preprint{GEF--TH/2006--4}

\title{\Large{{\bf Analysis of systematic errors in the calculation
of renormalization constants of the topological susceptibility on
the lattice   }}}
 \author{B. All\'es}
 \email{alles@df.unipi.it}
 \affiliation{INFN, Sezione di Pisa, Pisa, Italy}
 \author{M. D'Elia}
 \email{delia@ge.infn.it}
 \affiliation{Dipartimento di Fisica, Universit\`a di Genova and
INFN, Genoa, Italy}
 \author{A. Di Giacomo}
 \email{adriano.digiacomo@df.unipi.it}
 \affiliation{Dipartimento di Fisica, Universit\`a di Pisa and
INFN, Pisa, Italy}
 \author{C. Pica}
 \email{pica@df.unipi.it}
 \affiliation{Dipartimento di Fisica, Universit\`a di Pisa and
INFN, Pisa, Italy}

\pacs{11.15.Ha, 11.10.Gh}

\vfill
\begin{abstract}
A Ginsparg--Wilson based calibration of the topological charge
is used to calculate the renormalization constants which appear
in the field--theoretical determination of the topological
susceptibility on the lattice. A systematic comparison is made
with calculations based on cooling.
The two methods agree within present statistical errors (3\%--4\%).
We also discuss the independence of the multiplicative
renormalization constant $Z$ from the background topological
charge used to determine it.
\end{abstract}

\vfill

\maketitle

\section{Introduction}

Topology in QCD plays a relevant r\^{o}le in understanding
several low energy properties of the theory. One of the quantities
that have a direct interest in phenomenology is the topological
susceptibility~$\chi$ which is defined as the correlator at
zero momentum of two topological charge density operators $Q(x)$.
In particular the value of $\chi$ in the pure gauge theory
is interpreted in terms of the mass of the
singlet pseudoscalar meson~\cite{witten,veneziano}.

The lattice is an excellent tool to calculate such dimensionful
observables and specifically the lattice determination of $\chi$
in the pure gauge theory is in good agreement with phenomenological
expectations~\cite{alles,lucini,giusti}.

The calculation of $\chi$ or any other topology--related quantity
on the lattice requires a regularization $Q_L(x)$
of the topological charge density operator.
The formal na\"{\i}ve continuum limit must satisfy
$Q_L(x){\buildrel a\rightarrow 0 \over \longrightarrow}a^4Q(x)$,
($a$ is the lattice spacing).

However in general $Q_L(x)$ does not meet the continuum Ward
identities~\cite{crewther}. Consequently the lattice definition of
the topological susceptibility $\chi_L\equiv\langle Q_L^2\rangle/V$
($V$ is the lattice volume and $Q_L\equiv\sum_x Q_L(x)$)
need not coincide with the physical
continuum expression $\chi$. The two quantities are related
by~\cite{nupha329,haris}
\begin{equation}
\chi_L = Z^2 a^4 \chi + M\, ,
\label{chiL}
\end{equation}
where $Z$ and $M$ are renormalization constants
which for evident reasons are called
multiplicative and additive respectively.
In order to extract $\chi$ from the lattice data of $\chi_L$
one must know the values of $Z$ and $M$.

Let us outline the origin of the two renormalization constants.
Any matrix element which contains $n$ insertions of the topological charge
operator can be calculated either in the continuum (with an adequate
regularization) or on the lattice. In general the two calculations
match only after the inclusion of a multiplicative renormalization
constant $Z$~\cite{haris,vicari} for each insertion of $Q(x)$.
In a formal writing, $Q_L(x)=Z a^4 Q(x)$ where $Z$ is a finite
renormalization constant.
In the theory with fermions the topological charge mixes with
other operators related to the axial anomaly~\cite{espriu}. This mixing
induces a correction to the above described multiplicative renormalization
and consequently also to Eq.(\ref{chiL}).
Such a correction is however rather small~\cite{alles5} and it is usually
neglected consistently with the large statistical errors from a typical
numerical simulation.

If $n\geq 2$ the above matrix element includes further divergences which
are the origin of the additive term $M$. In fact the expression for
$\chi$ contains the product of two
topological charges at the same spacetime point. The operator expansion of
this product at short distances contains mixings with operators
that share the quantum numbers of $\chi$. On the other hand
it is known that the correlation function of two topological
charge operators at nonzero distance is
negative~\cite{osterwalder,seiler},
$\langle Q(x) Q(0)\rangle < 0$ for $x\not= 0$. This inequality
also holds on the lattice for any definition of
$Q_L(x)$~\cite{menotti,kirchner,horvath,ilgenfritz}. Since $\chi$
is a positive quantity, part of the contact terms must be
included in the physical definition of $\chi$. The rest of
the terms, if any, must be subtracted and they are $M$.

A prescription is necessary to calculate $M$. Due to the
expression
$\chi={\rm d}^2\ln Z(\theta)/{\rm d}\theta^2\vert_{\theta=0}$ where
$Z(\theta)$ is the partition function of the gauge theory
with a theta term, we know that $\chi$ vanishes within the
zero topological charge sector. We then adopt the
following definition for $M$: it is the value of
$\chi_L$ in the sector of zero topological charge,
$M\equiv\chi_L\vert_{q=0}$ where $q$ is the value of the
total topological charge of a configuration (as determined
by cooling or other means, see later).

A nonperturbative method to calculate $Z$ and $M$ has been
developed in Refs.~\cite{vicari2,gunduc}. The method will be
described in Section~2. It has been used in several calculations
of topological properties in QCD and other theories. 
Various tests and studies of efficiency have also been worked out
in the past. In Section~3 we will present the main contribution of the
paper: a study of systematic errors that may affect the calculation
of the renormalization constants and a comparison of results when
they are obtained by using different methods for calibrating the
background topological sector. Some conclusive comments
are given in Section~4.


\section{The nonperturbative calculation of $M$ and $Z$}

In Refs.~\cite{vicari2,gunduc} a technique, called ``heating method'',
to calculate the renormalization constants in Eq.(\ref{chiL}) was
put forward. 
The idea behind the heating method is that the UV fluctuations in
$Q_L(x)$, which are the ultimate cause for renormalizations,
are effectively decoupled from the background topological
signal so that, starting from a classical configuration of fixed
topological content, it is possible, by applying
a few updating (heating) steps at the 
corresponding value of the lattice bare coupling constant $\beta$,
to thermalize the UV fluctuations without altering the
background topological content. This result is favoured by
the fact that topological modes have very large autocorrelation
times, as compared to other non--topological observables
(this autocorrelation time is particularly long
in the case of full QCD~\cite{vicari3,lippert} and also in the
case of a large number of colours~\cite{ddpp}).

One can thus create samples of configurations within a given
topological sector $q$ with the UV fluctuations thermalized.
If $q\not=0$ then the measurement of $Q_L$ on such a sample leads
to the lattice value of the background topological charge $q_L$.
As described above~\cite{haris,vicari}, this result must be
renormalized to match the continuum value $q$,
\begin{equation}
q_L\equiv\langle Q_L\rangle\vert_q = Z q\, ,
\label{qLZq}
\end{equation}
where the subscript $\vert_q$ indicates that the thermalization
is achieved within the sector of charge $q$. Therefore, knowing
$q$ from the classical configuration and determining $q_L$ from
the measurement of $Q_L$ on the sample, $Z$ can be extracted.

If $q=0$ then measuring $Q_L^2/V$ on the sample leads to the
value $\chi_L\vert_{q=0}$ which is precisely $M$, as indicated
above.

$M$ can also be calculated on samples with nonvanishing topological
background $q$. Following Eq.(\ref{chiL}) and Eq.(\ref{qLZq}) the additive
constant is extracted in this case by using the relation 
\begin{equation}
 M= \frac{1}{V} \left(\langle Q_L^2\rangle\vert_q - 
                     \left(\langle Q_L\rangle\vert_q\right)^2\right)
\end{equation}
and leads to the same results~\cite{vicari2,delia}.

A sample of configurations belonging to the sector of total topological
charge $q$ is obtained in the following way. One starts from a classical
configuration with topological charge $q$. It can be easily obtained either
by using cooled configurations where the energy and the topological
charge correspond to the presence of one instanton
(if $q=1$)\footnote{Alternatively one can also approximate a~BPST
instanton~\cite{bpst} on the discrete lattice~\cite{fox}. The same
procedures can be applied for $q>1$.}
or by setting all gauge links to unity (if $q=0$).
Then a few updating steps are applied and the proper operator
(either $Q_L$ or $Q_L^2/V$) is measured at each step.
Moreover at each step the background topological
charge is checked by cooling\footnote{Different variants of the
cooling procedure lead to identical results~\cite{dino}.}~\cite{teper1}
in order to verify that the configuration still lies in the
sector of charge $q$~\cite{papafarchioni}. When the result of the
measurement stabilizes (data display a plateau), while $q$ stays fixed, we
consider that the UV fluctuations
are thermalized and $\langle Q_L\rangle\vert_{q=1}$ and
$\langle Q_L^2/V\rangle\vert_{q=0}$ yield $Z$ and $M$ respectively.

When the cooling applied to a configuration reveals that its background
topological charge is no longer equal to $q$ then the configuration
is discarded from the sample. However it might also happen the following
event: a possible new instanton or anti--instanton created by the
various updating steps
might evade the cooling probe because the very cooling procedure
could destroy it (this may happen especially when the instanton
spans a few lattice spacings). In this case we would include in the
sample a configuration which actually does not belong to the sector
of topological charge $q$. Such an event would obviously distort
the measurement of any of the above topological operators. For
example, it yields an
overestimation of $\langle Q_L^2\rangle\vert_{q=0}$ because any added
instanton or anti--instanton only increments the value of the square. On the
other hand since the theory predilects the sector of vanishing
topological charge, the updating steps during the calculation of
$Z$ tend to bring the
configuration to that sector either by destroying the original
instanton or by creating from scratch an anti--instanton.
As a consequence the value of $\langle Q_L\rangle\vert_{q=1}$
becomes underestimated because on average the
expectation value of $Q_L$ in the sector $q=0$ is less than
in the sector with $q=1$.

In the past we have always been aware of that potential problem and
in this paper we present a study where the background topological
sector is calibrated by another method in order to compare the results
and detect any difference in the form of a systematic error. The new
method is the counting of zero modes by using Ginsparg--Wilson
based operators~\cite{ginspargwilson} and it will be described in
the next Section.


\section{A study of systematic errors}

 Following the lines described in the above Section, we have
calculated the values of the renormalization constants in Eq.(\ref{chiL})
for the 1--smeared topological charge operator defined as~\cite{viejo}
\begin{equation}
 Q_L(x) = -\frac{1}{2^9 \pi^2}\sum_{\mu\nu\rho\sigma=\pm 1}^{\pm 4}
 \widetilde{\epsilon}_{\mu\nu\rho\sigma} {\rm Tr}\left\{\Pi_{\mu\nu}(x)
 \Pi_{\rho\sigma}(x)\right\}\,,
\label{Qlattice}
\end{equation}
where all link matrices have been substituted by 1--smeared
links~\cite{cristo} (the smearing parameter was $c=0.90$).
The corresponding renormalization constants will be called $M^{(1)}$ and
$Z^{(1)}$ to denote the level of smearing. In the above expression
$\Pi_{\mu\nu}(x)$ is the plaquette in the $\mu-\nu$ plane with
the four corners at $x$, $x+\widehat{\mu}$,
$x+\widehat{\mu}+\widehat{\nu}$, 
$x+\widehat{\nu}$ (counter--clockwise path).
Links pointing to negative directions mean
$U_{-\mu}(x)\equiv U_\mu^\dagger(x-\widehat{\mu})$.
The generalized completely
antisymmetric tensor is defined by $\widetilde{\epsilon}_{1234}=1$ and 
$\widetilde{\epsilon}_{(-\mu)\nu\rho\sigma}=
-\widetilde{\epsilon}_{\mu\nu\rho\sigma}$. The calculation was
performed for the Yang--Mills theory with SU(3) gauge group and Wilson
action~\cite{waction}. The lattice
size was $12^4$ and the bare coupling $\beta=6$.

The practical procedure was the following: starting from a
classical configuration with the adequate topological background
content ($q=0$ for $M^{(1)}$ or $q=1$ for $Z^{(1)}$), 80 heat--bath (HB)
steps were applied, each step consisting of three
Cabibbo--Marinari~\cite{cabibbom} hits, one for each 
SU(2) subgroup and Kennedy--Pendleton algorithm~\cite{kennedyp}
for refreshing the dynamical variables.
The operator ($(Q_L)^2/V$ for $M^{(1)}$ or $Q_L$ for $Z^{(1)}$) was measured
every 4 steps. This set of 20 measurements is called ``trajectory''.
A test of the topological sector was accomplished
after each measurement on a separate copy of the configuration.
We accepted only those measurements that were thermalized
within the corresponding topological sector. This condition
means that the
data must have stabilized to a plateau and that the test must have
revealed that the configuration lay within the correct topological
sector. The average over all accepted measurements
yielded $M^{(1)}$ or $Z^{(1)}$.

The test was performed by two different methods: either by a
traditional cooling~\cite{teper1}
or by counting fermionic zero modes~(CFZM).
This last method consists in calculating the net number
of zero modes, $n_+ - n_-$ by enumerating the
level crossings in the spectrum of the
Wilson--Dirac operator $D_W - a m$ as the fermion mass $m$
is varied~\cite{naraya,neuberg,edwards}. This method was
utilized in Refs.~\cite{edwards,deldebbio} to calculate
the topological charge.

The main advantage of the CFZM method against
cooling is that the first one does not need to
modify the configuration to which it is applied, hence the
topological content is never altered during the test. On the contrary
its main disadvantage is that its implementation is heavily
time--consuming.

Let us discuss in more detail the CFZM method.
We can stop counting crossings at any mass $am=am_{\rm stop}$
inside the allowed interval $am < 2$. Thus, in general the
measurement of the topological
charge will depend on $am_{\rm stop}$ because there can be level
crossings all along the interval of masses where the gap is
closed. This makes the zero mode counting method to look
ambiguous. However in Ref.~\cite{edwards} it is shown that
such a dependence is rather mild for $\beta\approx 6$ as long as
$am_{\rm stop}> 1.5$.
In the present work we have sought crossings up to three
different values for the stopping mass: $1.0$, $1.5$ and
$2.0$. In particular the last (and largest possible) value
looks particularly interesting because instantons representing
crossings close to $am=2.0$ span a size of a few lattice
spacings~\cite{edwards}, i.e.: they are the instantons that most
likely could evade the cooling test.

The search for crossings was realized by following the same
procedure of Ref.~\cite{deldebbio}. An accelerated conjugate gradient
algorithm~\cite{bunk} was employed to extract the lowest
eigenvalues of the Wilson--Dirac operator.

The topological charge after the cooling test
was required to be equal to 1 or 0 within a tolerance of $\delta$.
We usually chose $\delta=0.3$ although tests with other values
were performed (see later) proving that the results are very
robust against variability of $\delta$.

\begin{figure}[t]
\begin{center}
\includegraphics[angle=-90, width = 0.99\textwidth]{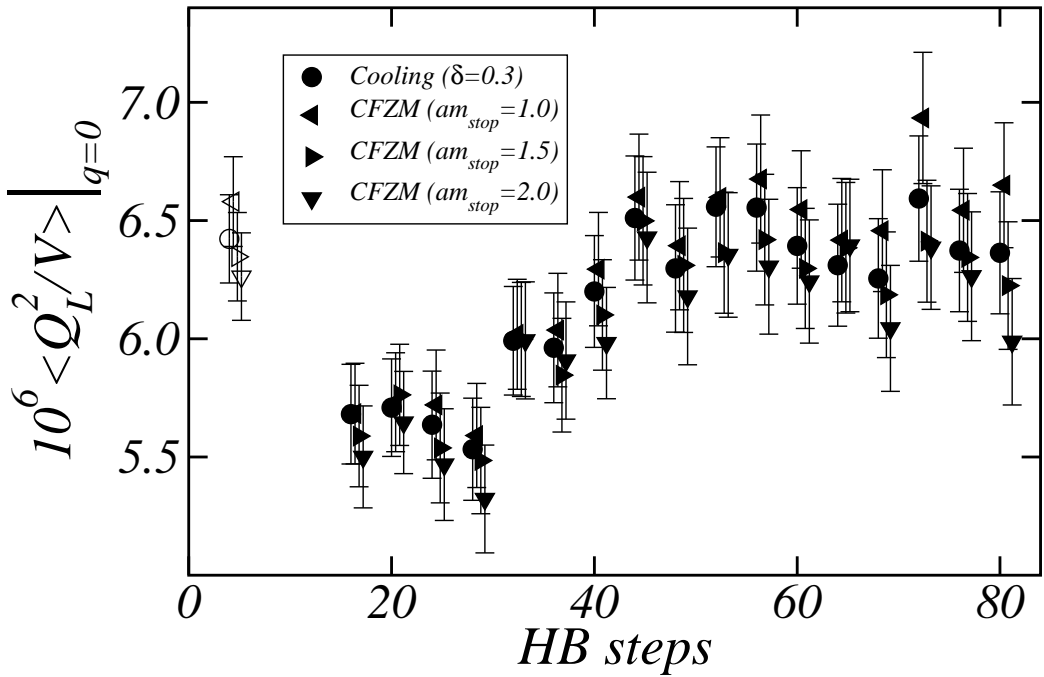}
\vskip -2cm
\caption{$\langle Q_L^2/V\rangle\vert_{q=0}$ as a function of the
HB step for the four calibration tests described in the text.
The results for $M^{(1)}$ have been placed on the left side and are
represented by white
symbols. Data corresponding to different calibration methods have been
shifted with respect to each other in order to render the Figure
clearer.}
\label{Mtotal}
\end{center}
\end{figure}

\subsection{Systematic errors on $M^{(1)}$}

The results for $M^{(1)}$ are shown in Fig.~\ref{Mtotal}. We have
calculated 1380 trajectories of 80 HB steps and measured
the operator $Q_L^2/V$ every 4 steps after checking that the background
charge was zero. For each measurement, in
the Figure we show the average over all the results. Actually we 
show the data after the 16$th$ step because measurements after
too few HB steps are irrelevant as the configuration
is surely not yet thermalized. A plateau seems to set in after
approximately 40 steps which indicates that thermalization
has been achieved. Then the value of $M^{(1)}$ can be extracted by
averaging over the data after the 40$th$ step.

The error bar was calculated in the following manner:
each trajectory was treated as a single data by averaging all accepted
(i.e.: all thermalized and correctly calibrated) points in it. Since separate
trajectories are prepared by independent Monte Carlo runs, this procedure
guarantees the absence of autocorrelations. Then the average and
the error are easily obtained from this set of 1380 data.
Furthermore, due to the fact that different trajectories can contain
a variable number of accepted points, each trajectory must be weighted
by a factor proportional to the number of accepted points in it.

\begin{figure}[t]
\begin{center}
\includegraphics[angle=-90, width = 0.99\textwidth]{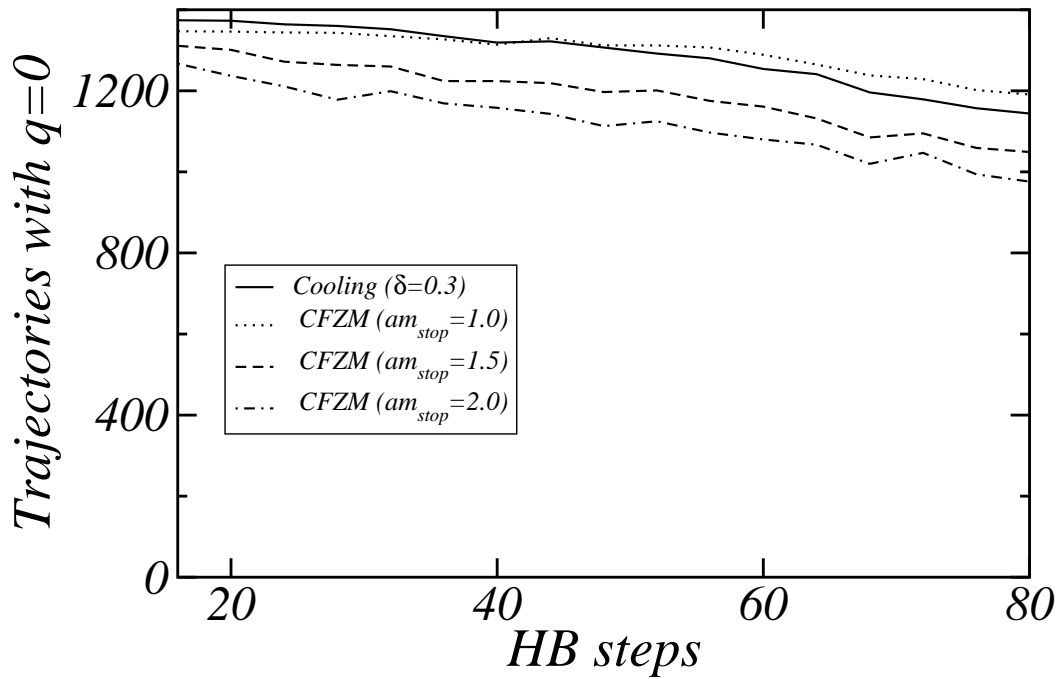}
\vskip -2cm
\caption{Counting of trajectories that still lie in the $q=0$ topological
sector as the number of HB steps increases for the four
calibration methods.}
\label{Mq0}
\end{center}
\end{figure}

If the number of accepted points in the trajectory $t$ is $n_t$ and
if $T_t$ is the average over all accepted measurements $m$ in
this trajectory,
\begin{equation}
 T_t = \frac{1}{n_t}\sum_{m\in t}
  \left( \frac{Q^2_L}{V}\right)_m
  \Bigg\vert_{\buildrel q=0 \over {\rm thermalized}}\; ,
\end{equation}
then
\begin{equation}
 M^{(1)} = \frac{\sum_t T_t n_t}{\sum_t n_t}\; .
\end{equation}

{}Fig.~\ref{Mtotal} displays the results for all four methods of calibration,
cooling with $\delta=0.3$ and CFZM with three different
values for the stopping mass.
The four white symbols on the left side of the Figure are the corresponding
results for $M^{(1)}$. They become smaller as the value of $am_{\rm stop}$
is incremented. This effect is possibly an indication of a systematic
error which however has little effect in the calculation since all results
look compatible with each other within errors.
It must be stressed that the statistical errors, represented by the
error bars in the Figure, amount to about 3\% which is rather small.

In Fig.~\ref{Mq0} we show, for each measurement, the number of trajectories
for which the calibration test gave an acceptable result, $q=0$.
The plot is shown for all measurements, thermalized or not, from the 20$th$
HB step onward. The maximum
possible number is obviously 1380 and it falls off as the amount of
HB steps is increased. The decrease is steeper for the CFZM
with larger stopping mass. In the latest step and for the CFZM
with $am_{\rm stop}=2$ about 30\% of all trajectories have varied the
topological sector, while for the cooling method the analogous percentage
is about 17\%. Such a difference still allows to obtain results for
$M^{(1)}$ that are compatible within (small) errors, as shown by the
white symbols in Fig.~\ref{Mtotal}.

\begin{figure}[t]
\begin{center}
\includegraphics[angle=-90, width = 0.99\textwidth]{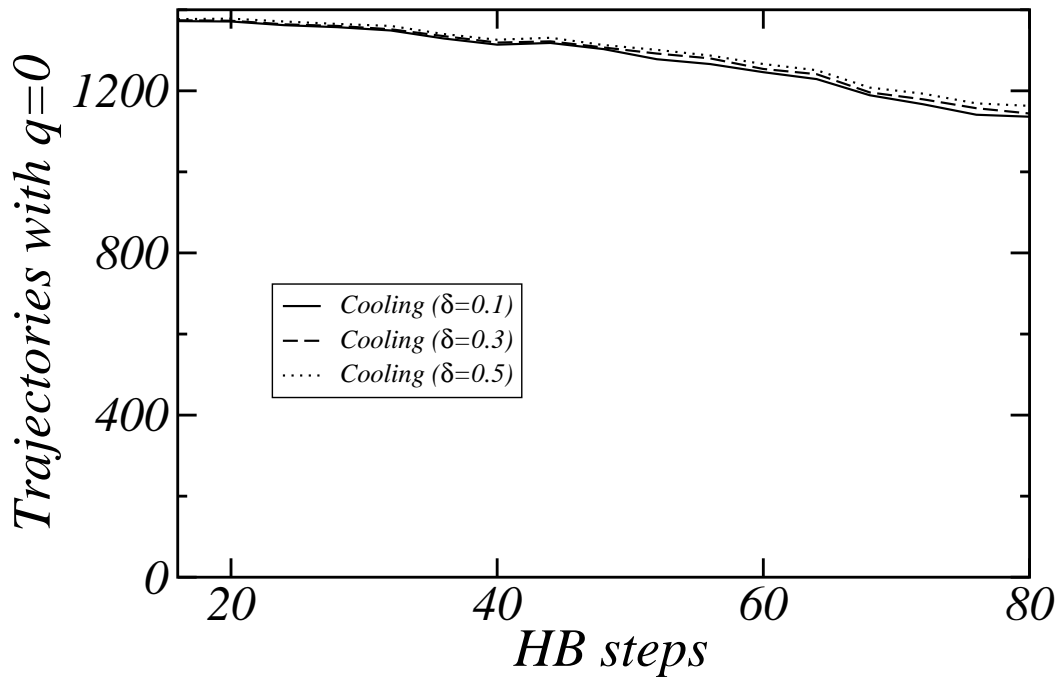}
\vskip -2cm
\caption{Counting of trajectories that still lie in the $q=0$ topological
sector as the number of HB steps increases for the cooling method with
three different values of $\delta$.}
\label{Mcoolq0}
\end{center}
\end{figure}

The values of $M^{(1)}$ obtained by using cooling with varying $\delta$ are
indistinguishable from each other (white circle in Fig.~\ref{Mtotal})
for $\delta$ ranging from 0.1 to 0.5.
In Fig.~\ref{Mcoolq0} three tolerance parameters for cooling are compared
indicating that the three tests are almost completely equivalent. Again the
maximum possible number of trajectories with the right background topological
charge is 1380 and it diminishes as the amount of HB steps is increased.
This number became 1136 for $\delta=0.1$ at the latest step.

\begin{figure}[t]
\begin{center}
\includegraphics[angle=-90, width = 0.99\textwidth]{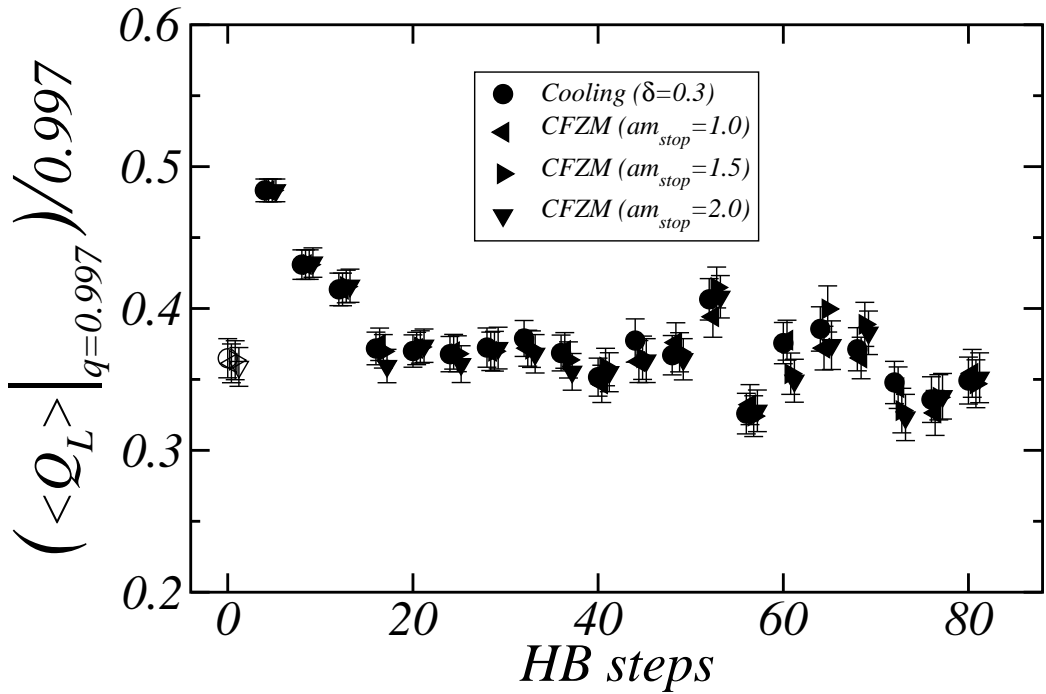}
\vskip -2cm
\caption{Averages of $\left(\langle Q_L\rangle\vert_{q}\right)/q$
as a function of the HB step for the four calibration methods.
An instanton with charge $q=0.997$
was used. The values of $Z^{(1)}$ correspond to the height of the plateau
and they are indicated on the left hand side of the plot with white
symbols. Data corresponding to different calibration methods have been
slightly shifted for clarity.}
\label{Zhator}
\end{center}
\end{figure}

\subsection{Systematic errors on $Z^{(1)}$}

{}Fig.~\ref{Zhator} displays the analogous study of Fig.~\ref{Mtotal}
for the calculation
of $Z^{(1)}$. An instanton with topological charge (as measured with $Q_L$
after cooling) $q=0.997$ was employed.
The procedure for the calculation of $Z^{(1)}$ resembles very much that
of the $M^{(1)}$. We prepared 840 independent trajectories and again the
calibration was performed by four different tests, as indicated in
Fig.~\ref{Zhator}. The average per trajectory is $T_t$,
\begin{equation}
 T_t = \frac{1}{n_t}\sum_{m\in t}
  \left( Q_L\right)_m
  \Bigg\vert_{\buildrel q=0.997 \over {\rm thermalized}}\; ,
\end{equation}
and the result for $Z^{(1)}$ is
\begin{equation}
 Z^{(1)} = \frac{1}{0.997}\,\frac{\sum_t T_t n_t}{\sum_t n_t}\; .
\label{z997}
\end{equation}
A plateau sets in at about the 20$th$ HB step. Figures similar
to Fig.~\ref{Mq0} and~\ref{Mcoolq0} are obtained with analogous conclusions.
Again the coincident results (white symbols in Fig.~\ref{Zhator})
indicate that all calibration methods are equivalent in such a
way that any systematic error implicit in our method has negligible
consequences within our precision (statistical errors in Fig.~\ref{Zhator}
are about 4\%).

\begin{table}
\caption{$Z^{(1)}$ for the 1--smeared operator as a function of $\beta$ for
several values of the background topological charge~$q$. Instantons
with charge $q\approx -1$, $\approx +1$ and $\approx +2$ are used on
a lattice volume $16^4$.}
\vspace{0.25cm}
\begin{tabular}{|c|c|c|c|c|c|}
\hline \rule{0pt}{4ex}
$\; \beta\; $    &
$\; q=+0.990\; $ &
$\; q=-0.988\; $ &
$\; q=+0.955\; $ &
$\; q=-0.902\; $ &
$\; q=+1.941\; $ \\
\hline \rule{0pt}{3.5ex} 6.00 & 0.373(20) & 0.383(15) &
                                0.370(20) & 0.365(20) & 0.390(12) \\
\hline \rule{0pt}{3.5ex} 6.20 & 0.432(6)  & --  & -- & -- & 0.441(5)  \\
\hline \rule{0pt}{3.5ex} 6.50 & 0.503(11) & --  & -- & -- & 0.506(5)  \\
\hline
\end{tabular}
\label{z1}
\end{table}
\begin{table}
\caption{$Z^{(2)}$ for the 2--smeared operator as a function of $\beta$ for
several values of the background topological charge~$q$. Instantons
with charge $q\approx -1$, $\approx +1$ and $\approx +2$ are used on
a lattice volume $16^4$.}
\vspace{0.25cm}
\begin{tabular}{|c|c|c|c|c|c|}
\hline \rule{0pt}{4ex}
$\; \beta\; $    &
$\; q=+0.990\; $ &
$\; q=-0.988\; $ &
$\; q=+0.955\; $ &
$\; q=-0.902\; $ &
$\; q=+1.941\; $ \\
\hline \rule{0pt}{3.5ex} 6.00 & 0.500(12) & 0.503(13) &
                                0.495(15) & 0.487(20) & 0.499(16) \\
\hline \rule{0pt}{3.5ex} 6.20 & 0.560(6)  & -- & -- & -- & 0.569(4)  \\
\hline \rule{0pt}{3.5ex} 6.50 & 0.620(10) & -- & -- & -- & 0.631(3)  \\
\hline
\end{tabular}
\label{z2}
\end{table}

\begin{figure}[t]
\begin{center}
\includegraphics[angle=-90, width = 0.99\textwidth]{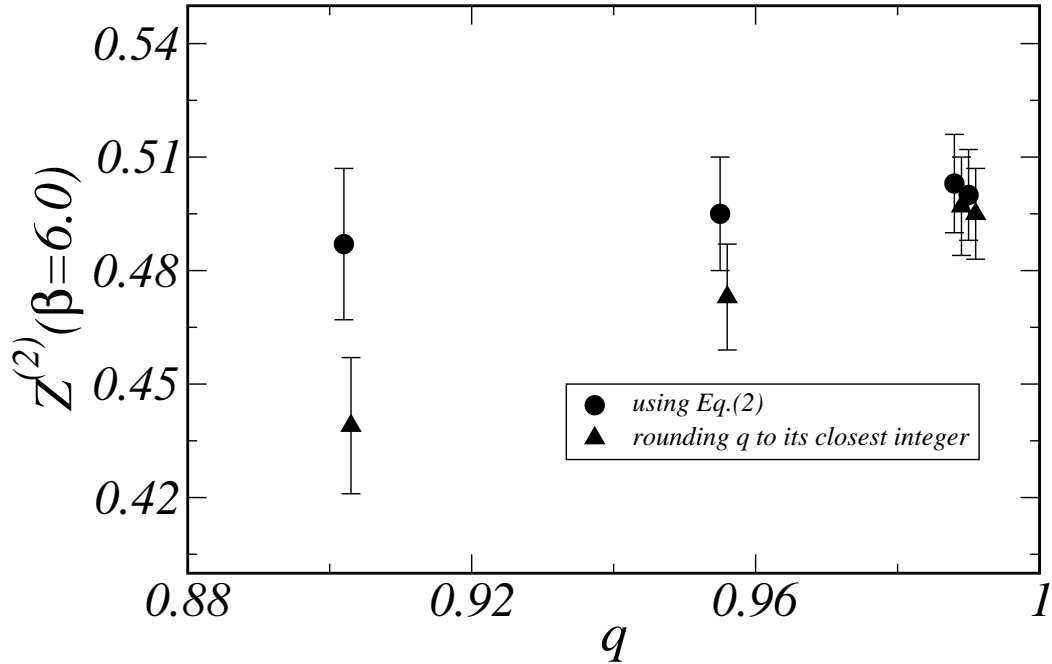}
\vskip -2cm
\caption{$Z^{(2)}(\beta=6.0)$ as obtained by using four several initial
classical instantons (with $|q|=$0.902, 0.955, 0.988 and 0.0990).
Squares stand for the results obtained by applying Eq.(\ref{qLZq})
and triangles by applying the same equation after rounding the
value of $q$ to its closest integer.}
\label{Z2smear}
\end{center}
\end{figure}

It is well--known that the
total topological charge of isolated classical instantons in general
is not equal to integer numbers when it is calculated with the
operator of Eq.(\ref{Qlattice}). The difference between the value
of the charge and the closest corresponding integer ($\vert 1-q\vert$ in
our case) becomes
negligible when the inequalities $a \ll \rho \ll La$ hold ($\rho$
being the instanton size and $L$ the lattice size).
A simple calculation shows that the value of $q$ for a discretized
instanton in a volume $L^4$ is given by
\begin{equation}
 q\approx 1 - 3 \left(\frac{\rho}{La}\right)^4\; ,
\label{approxQ}
\end{equation}
while the discretization error is totally negligible if $a\lesssim\rho/3$.
The ratio $a/\rho$ can be estimated by looking at the action density
distribution. We expect that the infrared effect
described in Eq.(\ref{approxQ}) cancels out if we divide the
charge after heating by the initial value $q$ as indicated in
Eq.(\ref{qLZq}). This fact was carefully checked in~\cite{becca}
for the 2D $O(3)$ nonlinear sigma model. It was also used in Eq.(\ref{z997})
although in that case and within our errors, the $q=0.997$ in the
denominator is indistinguishable from 1. In the present study
we have verified it for our gauge theory: in Table~\ref{z1} and~\ref{z2}
the multiplicative constant is calculated at several values of
the gauge bare coupling $\beta$
on a $16^4$ volume starting from various initial instantons
for the 1-- and 2--smeared operators
(they are constructed as in Eq.(\ref{Qlattice}) after substituting
all links with 1-- and 2--smeared links respectively~\cite{cristo}).
A number of trajectories ranging from 200 to 500 were used. Notice
that within errors the values for $Z^{(1)}$ and $Z^{(2)}$
display a dependence on $\beta$ but not on $q$ as long as data
are divided by the initial value $q$, as described above
following Eq.(\ref{qLZq}).
If instead data were divided by the integer closest to $q$ then
the results for $Z^{(i)}$, $(i=1,2)$ would display a
fake dependence also on $q$.
This fact is seen in Fig.~\ref{Z2smear} where the values of
$Z^{(2)}$ are displayed as a function of $q$ after dividing by
$q$ (squares) or by the closest integer to $q$ (triangles):
only the squares show constancy with respect to $q$.
Within our statistical errors the systematic error is negligible
down to $q\approx 0.9$. Possibly if the statistics were increased
this limit could grow. In any case to calculate $Z$ it is better
to make use of instantons with $q$ as close as possible to an
integer.

\section{Conclusions}

We have studied two possible sources of systematic
errors in the nonperturbative determination of the renormalization
constants for the evaluation of the topological susceptibility
on the lattice, $Z$ and $M$ (see Eq.(\ref{chiL})):

\vskip 2mm

\begin{itemize}

\item {\it i)} The cooling test of the background topological charge
along the heating process could yield a wrong information since
the cooling procedure could delete
the unwanted instanton or anti--instanton created by the heating steps.
An independent check of the cooling test has been performed
by applying calibration methods based on the counting
of fermionic zero modes. No sizeable systematic
effects have been observed within our (rather small)
statistical errors (3\% for $M$ and 4\% for $Z$).

\item {\it ii)} In the calculation of $Z$ for the operator $Q_L$ of
Eq.(\ref{Qlattice}) the simulation must be
started from a configuration with a topological charge $q$ different
from zero. In the infinite volume limit $q$ takes on integer values, however on
the lattice, the determination of $q$ by using the operator
in Eq.(\ref{Qlattice}), leads to results which in general are close to
but not strictly equal to integers.
We have argued that this is a potential source of
error that can affect the calculation of $Z$. However it can be
kept under control if one uses Eq.(\ref{qLZq}) to extract $Z$
with $q$ not rounded to its closest integer.

\end{itemize}

\vskip 2mm

In conclusion one can safely use the field--theoretical method
to study topology on the lattice since any possible systematic error
of the method is well under control. Moreover the method is much
less demanding in computer time than the Ginsparg--Wilson based method.

The APEmille facility in Pisa was used for part of the runs.

 \section*{Acknowledgements}
We thank M.I.U.R. for financial support, Project number
2004021808\_004.

\end{document}